# Polarization Analysis and Probable Origin of Bright Noctilucent Clouds with Large Particles in June 2018

Ugolnikov O.S.[1], Maslov I.A.[1,2]

[1]Space Research Institute, Russian Academy of Sciences,
84/32 Profsoyuznaya st., Moscow, 117997, Russia
[2]Moscow State University, Sternberg Astronomical Institute,
13 Universitetsky pr., Moscow, 119234, Russia

E-mail : ougolnikov@gmail.com, imaslov@iki.rssi.ru

A sequence of very bright noctilucent clouds (NLC) was observed in central Russia during the last decade of June, 2018. It followed the meteorite impact 300 km southwards from Moscow in the morning of June 21. Polarization measurements of NLC in a wide range of scattering angles allowed finding the effective size of ice particles forming the clouds. These estimations together with satellite data on gravity waves, temperature, and water vapor mixing ratio during these days help to understand the basic causes of the observed NLC brightness and particle size anomaly.

## 1. Introduction

The optical phenomenon of noctilucent clouds has been observed for less than 150 years. For the first time, they occurred in 1885 and were observed in many countries of central and northern Europe (Leslie, 1885). The event of highest clouds in the atmosphere (at an altitude of above 80 km) was associated with the Krakatoa volcano eruption in August 1883. The hypothesis of water ice content in NLC was suggested in the early XX century (Wegener, 1912) and was finally confirmed a century later (Hervig et al., 2001). Volcanic dust can play the role of condensation nuclei if the eruption is strong enough to throw a large amount of dust into the mesosphere.

However, NLC were also observed long after the Krakatoa epoch: the most remarkable event happened in the first days of July 1908. Immediately after the Tunguska airburst in the atmosphere above Siberia in June 30, the nights became very bright above vast territories in Europe and Asia westward from the impact location. This was probably caused not by strong zonal winds in the upper atmosphere but by the comet nature of the Tunguska body, which was possibly a fragment of the well-known Encke comet. NLC particles could nucleate on the dust of Tunguska body's coma entered the atmosphere far from impact location. Now we know that space bodies form tiny "meteor smoke" nanometer-sized particles (Hunten et al., 1980), which were detected by Havnes et al. (1996).

Other possible nuclei for NLC particles are hydrate ions in the mesosphere (Witt, 1969). The model of NLC particles nucleation and evolution was initially performed by Turco et al. (1982) and then improved by Rapp and Thomas (2006). The process strongly depends on the physical state of this atmospheric layer, basically, on the water vapor content and temperature. For the upper mesosphere pressure and typical volume mixing ratio (VMR) of $H_2O$ (about several ppmv), the frost temperature is between 140K and 150K. Such low temperatures are reached only in polar regions in summer. The models show that the particle size increases as it moves downwards in the cold upper mesospheric layer, reaching 50–100 nm at 80–82 km. Lidar, spaceborne and rocketborne measurements (see reviews of results in (Kokhanovsky, 2005; Baumgarten et al., 2008)) provided the typical values of the effective particle radius (the radius of monodisperse particle ensemble with the same optical properties) of 50–60 nm, the median radius of lognormal distribution (von Savigny and Burrows, 2007) is half the size of the former, about 30 nm. Note that the light scattering efficiency of very small particles increases with their radius $r$ as $r^6$, and we analyze the largest (and usually the lowest) fraction of particles in optical observations. Particle size, NLC brightness and occurrence increase in colder conditions, when the layer with $T<T_{FROST}$ is thick (Feofilov and Petelina, 2010). This is the reason of the mean particle size increase as we move polewards (von Savigny and Burrows, 2007).

Bright clouds around the pole are unseen from the Earth due to polar day conditions; however, at latitudes 50°–60° NLC can be seen during summer twilight, so this latitude belt is optimal for ground-based NLC observations. Polarization study using all-sky camera can be effective for particle size retrieval if NLC are registered both northwards and southwards from the zenith and available data is expanded to scattering angles $\theta>90°$. This analysis was done by Ugolnikov et al. (2016); the same techniques are used in this paper.

Another important NLC formation factor is the propagation of gravity waves through the mesosphere (Witt, 1962; Fritts et al., 1993). Their amplitude increases in the upper mesosphere, which can cause significant temperature changes and wavelike structures of NLC. As the wave moves upwards, the altitude of cold layers and, thus, the size of NLC particles change (von Savigny et al., 2005; Megner et al., 2016; Rusch et al., 2017).

The importance of a certain factor for NLC occurrence and characteristics can be studied by correlation analysis (Feofilov and Petelina, 2010) or if this factor changes rapidly or reaches an unusual value. The influence of the meteoric dust inflow was obvious after the Tunguska event. The occurrence of bright NLC occasionally shows correlations with major meteor showers. The strongest summer shower is Perseids with a maximum on August 12–13. During those days, the mesosphere is cold enough to form clouds near the Polar Circle; bright NLC were registered by a polarization camera in Apatity (Russia, 67.5°N) during the first summer observation night on August 15, 2015, when the Perseids were still active (Ugolnikov et al., 2016); a year later, even brighter NLC were observed at Lovozero station (68°N), straight after the shower maximum (August 12, 2006). The results of measurements of particle size made by an RGB all-sky camera (Ugolnikov et al., 2017) were normal (the effective size was about 55 nm), and the brightness of clouds could be explained by a greater number of particles.

The impact of a large meteor body significantly increases dust density in various atmospheric layers for a long time. The asteroid impact near Chelyabinsk, Russia, on February 15, 2013 caused the strongest airburst after the Tunguska event (Popova et al., 2013). The impact was followed by appearance of a dust belt in the upper stratosphere, which was observed for several months (Gorkavyi et al., 2013). Being driven by strong zonal winds, the dust had completed several revolutions around the Earth's axis, forming a ring at latitudes 50–60°N. Although it occurred during the warm mesosphere season, the wavelike structures visually similar to NLC were observed in Moscow during several twilights in late winter and early spring.

In the morning of June 21, 2018 (1h15m UT), a meteorite impact was registered in Russia (52.8°N, 38.1°E), about 300 km southwards from Moscow. It was the first meteorite event near the summer solstice in Russia after Tunguska. The explosion power was estimated at about 1% of the Chelyabinsk airburst. It was followed by bright noctilucent clouds observed in many locations in Russia and Europe until the end of June. The basic aim of this paper is to study the properties of particles and physical conditions in the mesosphere when NLC appeared. It will help to understand if there is any relation between these events and to find the cause of such a remarkable appearance of NLC during this period.

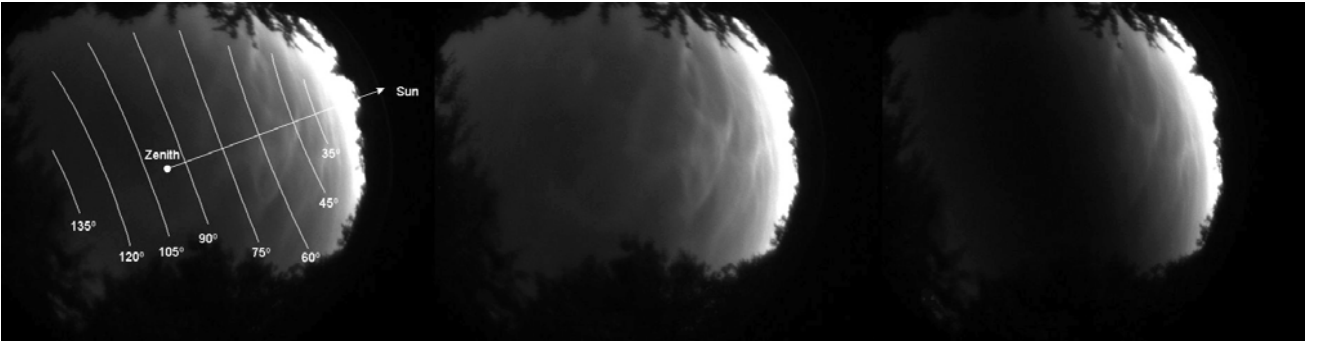

*Figure 1. Images of the sky with NLC with different polarization directions (120° one from another), evening of June, 25. Right image corresponds to the polarization direction parallel to solar vertical, thick black line in the sky shows the maximum of polarization degree. The lines of constant scattering angle are shown in left image.*

**2. Observations and particle size estimation**

Bright noctilucent clouds were observed near Moscow under clear sky conditions twice during late June of 2018, on the nights of day 25 (the maximum at around 20h UT, local evening) and 27 (23h UT, local morning). In both cases, the clouds extended southwards from the zenith to the scattering angles of about 120°. During those nights, the sky background was measured with a Wide-Angle Polarization Camera (WAPC, 55.2°N, 37.5°E, Ugolnikov and Maslov, 2013ab). A detailed description of the device is provided therein: it was a zenith-oriented camera with the field diameter of 140°, designed for intensity and polarization measurements in the spectral band with the effective wavelength of 540 nm and FWHM of 90 nm. The measurements were conducted from the sunset till the sunrise; night star images were used to adjust the camera's direction and geometric field parameters.

WAPC has been operating since 2011 and already registered a bright NLC field with visual properties similar to the present events during the night of July 5, 2015. This case was considered in detail by Ugolnikov et al. (2016); the same paper provides a description of the data processing procedure, sky background reduction, and NLC polarization retrieval. This procedure was replicated herein without any changes. Figure 1 shows NLC images in the evening of June, 25 in three polarization directions (120° from each other). As is seen, the cloud field is strongly polarized near the scattering angle of 90°; this is the property of tiny particles forming the NLC. However, the dependence of polarization (or second Stokes component) on the scattering angle can differ from the Rayleigh function, this is the basis of the particle size estimation. For the spectral band used herein and ice particle size less than 160 nm, this effect manifests in a shift of the maximum polarization to higher scattering angles rather than in a polarization decrease.

Figure 2 shows the dependence of the NLC field polarization on the scattering angle (or angular distance from the Sun): the polarization component transverse to the sunward direction is plotted for the nights of June 25 and 27. Theoretical dependencies for tiny spheres (Rayleigh) and monodisperse ice spheres with a radius of 100 nm are also plotted. Following Ugolnikov et al. (2016), we find most probable parameters of lognormal particle size distribution (mean radius $r$, distribution width $\sigma$) by minimizing the value:

$$\chi^2 = \frac{1}{\sqrt{2N}}\left(\sum_{n=1}^{N}\frac{(p(\theta_n)-p_T(r,\sigma,\theta_n))^2}{\sigma_M^2(\theta_n)} - N\right). \qquad (1)$$

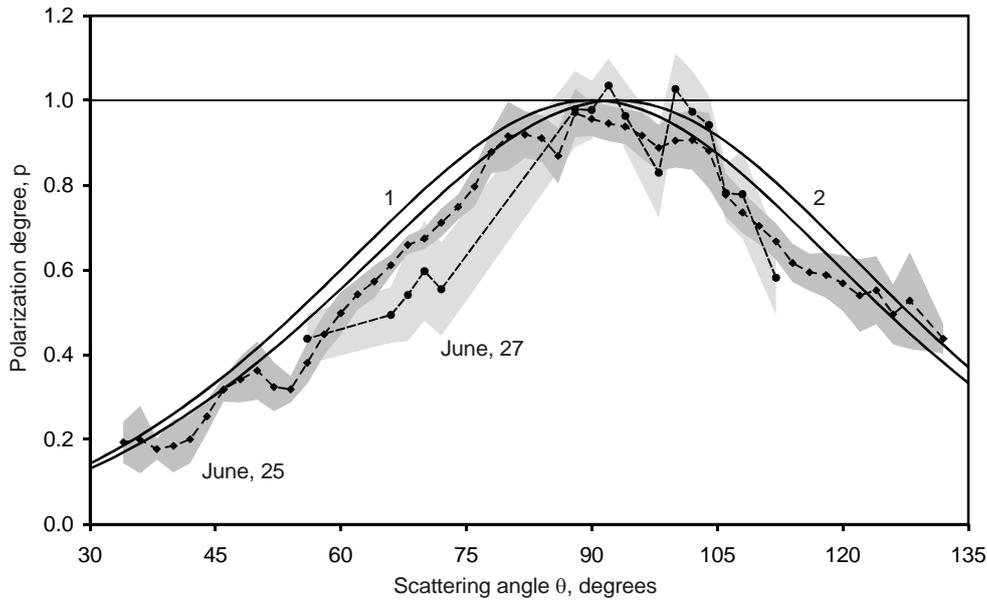

*Figure 2. Angular dependencies of polarization of light scattered by NLC particles. Model curve 1 corresponds to tiny (Rayleigh) particles, curve 2 corresponds to the particles with radius 100 nm.*

Here $p$ and $p_T$ mean measured and theoretical (Mie) values of polarization for scattering angle $\theta_n$, $\sigma_M$ is the accuracy of measurements, $N$ is the total number of scattering angles. Results for monodiserse particles ($\sigma=1.0$) and for $\sigma=1.4$ are listed in Table 1, the values for July, 5, 2015 are also provided for comparison. We see that the effective radius of ice (case $\sigma=1.0$) significantly exceeds a typical value of 50-60 nm received in many lidar, spaceborn and rocket experiments and polarimetry in 2015. Figure 3 shows the distribution of $\chi^2$ on $(r - \sigma)$ diagram for both observation dates. It is worthy of note that wide distributions of particle sizes are more probable. We can also add that if possible non-spherical particles with the axes ratio of up to 2 are taken into account, it does not change the estimated size by more than 10-15 nm.

The basic conclusion of this chapter is that bright NLC observed in July 2018, at least during WAPC observations near Moscow, are characterized by large particle sizes. The effective value of $r$ is about 100 nm; the mean particle size in the case of lognormal distribution (width 1.4, von Savigny and Burrows, 2007, also shown in Table 1) is about 60 nm. It is twice the normal value typically measured for NLC. Since tiny particles have rapid dependence of scattering intensity on size ($\sim r^6$), it is enough to explain NLC brightness and no increase of meteor dust particles possibly caused by meteorite impact is needed for this. However, one should also consider the physical properties of the upper mesosphere on the days of NLC occurrence to find possible reasons of such a large particle size. The next chapter deals with this problem.

| Date | Effective radius ($\sigma=1.0$), nm | Mean radius ($\sigma=1.4$), nm |
|---|---|---|
| July, 5, 2015 | $59^{+34}_{-59}$ | $27^{+15}_{-27}$ |
| June, 25, 2018 | $107^{+11}_{-16}$ | $64^{+9}_{-9}$ |
| June, 27, 2018 | $112^{+24}_{-74}$ | $74^{+16}_{-24}$ |

*Table 1. Retrieved values of effective particle radius (monodisperse model) and mean particle radius (lognormal distribution with $\sigma=1.4$) basing on polarization measurements of NLC.*

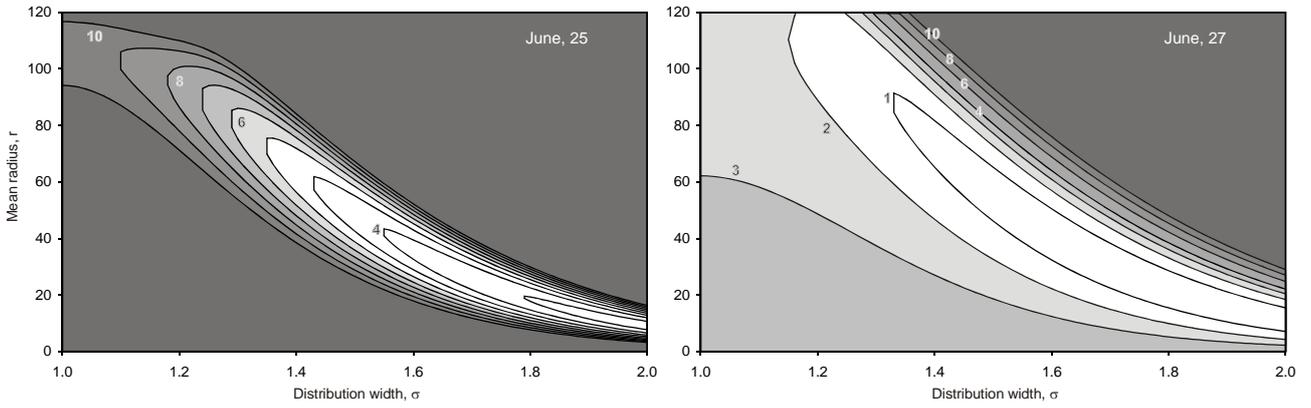

*Figure 3. Dependence of $\chi^2$ on the parameters of lognormal distribution of particle sizes. Monodisperse distribution corresponds to $\sigma=1.0$.*

## 3. Local physical conditions in the mesosphere

The basic factors defining the NLC occurrence and properties, except the meteoric dust, are temperature and the water vapor mixing ratio (Feofilov and Petelina, 2010; Christensen et al., 2016). Temperature strongly depends on gravity waves, differing by more than 20K at a half-wave distance. This phenomenon can be recorded using high-resolution temperature profiles provided by TIMED (*Thermosphere Ionosphere Mesosphere Energetics Dynamics*) / SABER (*Sounding of the Atmosphere Using Broadband Emission Radiometry*) satellite measurements (Russell et al., 1999). Figure 4 shows the profiles of all SABER scans between the 15th and 30th of June 2018 at locations at ±2° latitude and ±5° longitude from the observation place. These restrictions are defined by the typical horizontal length of gravity waves in the upper mesosphere, about 300 km (Rusch et al., 2017). As is seen, all profiles have wave-like structures. One of them refers to the NLC twilight on June 25; it was built just two hours after the NLC observation. It is the scan characterized by the lowest temperature at the layer of visual clouds, between 80 and 82 km. The value is as low as 125K, increasing above the frost level at higher altitudes. It seems to be exactly the case described by Rusch et al. (2017) when very large particles can be observed. It should also be noted that two other deep temperature minima in Figure 4 correspond to daytime and bad weather conditions respectively.

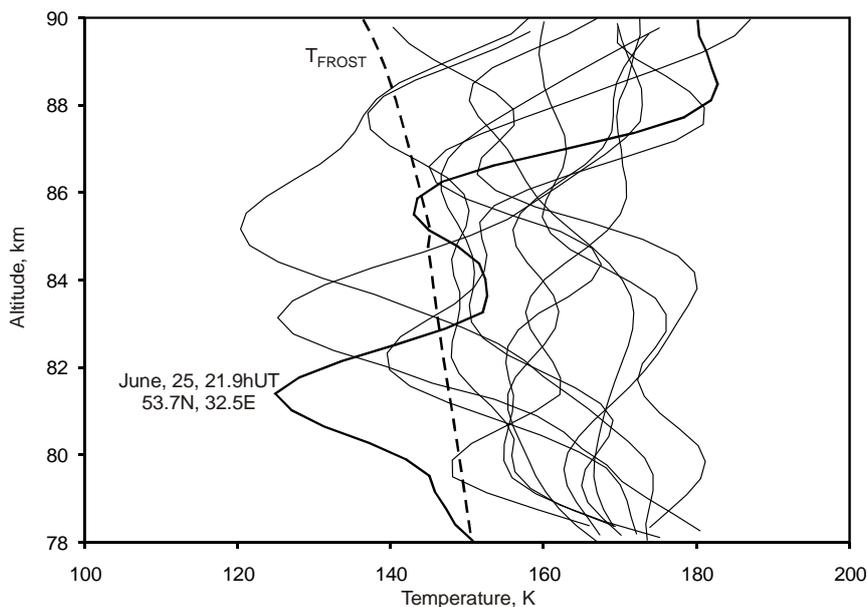

*Figure 4. SABER mesopause temperature scans for locations close to the observations site in June 18-30, 2018.*

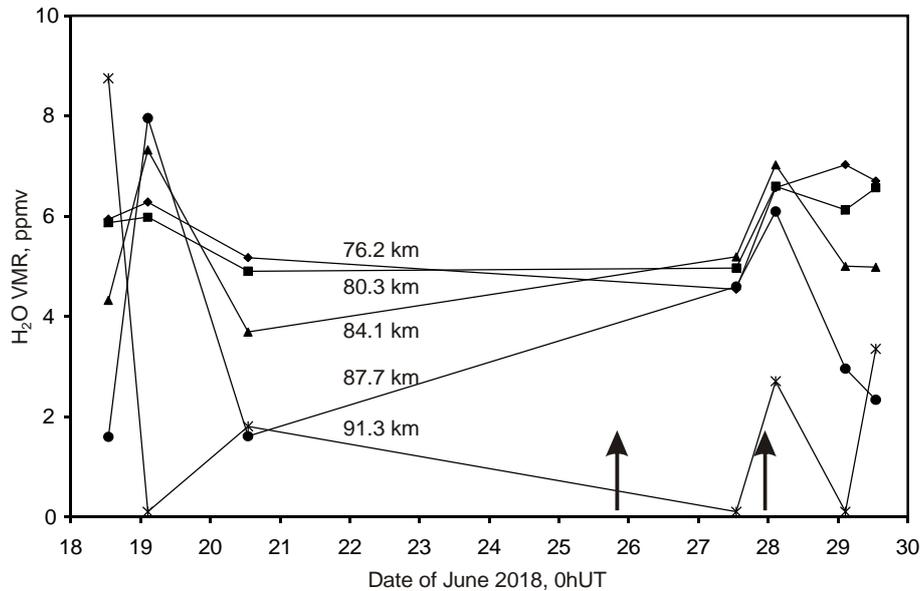

*Figure 5. MLS values of H₂O volume mixing ratio for nearby scans in June 18-30, 2018. The moments of NLC observation are shown by arrows.*

There is no nearby scan of SABER temperature in the second twilight of bright NLC appearance. However, the satellite data of EOS (Earth Observing System) Aura / MLS (Microwave Limb Sounder, EOS Team, 2011) on $H_2O$ mixing ratio is available for that day. Figure 5 indicates it for all MLS scans at a nearby latitude (54.7°N) and longitudes ±5° from the observation place. The typical $H_2O$ profile from this data and empirical formula (Murphy and Koop, 2005) are also used to plot the $T_{FROST}$ profile in Figure 4. One can see that the night of June 27 is characterized by a higher $H_2O$ mixing ratio in the whole layer between 80 and 90 km, which could also cause the bright NLC.

Finally, we can resume that ground-based polarization measurements and satellite data show no signs of the assumption that bright and long-observed noctilucent clouds in central Russia in June 2018 could be caused by a higher number of meteor dust particles and, thus, a higher number of ice cloud particles. What increased was the mean particle size, this was due to gravity waves and low temperature at the 80–82 km level and/or a high volume mixing ratio of water vapor.

## 4. Discussion and conclusion

The paper deals with the analysis of bright noctilucent clouds observed during several days in a row in central Russia in late June of 2018. They were observed in different locations during every night; some cases (including the evening of June 26 in the observation place) are missing from this analysis due to bad weather conditions.

This event directly followed the meteorite impact in a nearby location in the morning of June 21. The question of possible interconnection of these phenomena seems quite obvious, taking into account the "bright nights" in Europe after the Tunguska impact in 1908 and upper stratospheric aerosol observed after the Chelyabinsk meteorite airburst in 2013. Roughly assuming, by analogy with the Chelyabinsk event, that bright NLC in the night of June 27 caused by tiny dust fragments completed one revolution around the Earth's axis after the impact, we can find a realistic value of zonal wind in the upper mesosphere: 40 m/s.

Imagining this scenario, we can expect an increase in the number of meteor smoke particles and ice particles with usual size distribution. This pattern took place in August 2016 when bright NLC were observed at 68°N immediately after the maximum of the Perseids meteor shower (Ugolnikov et al., 2017). Large particles are less probable in this case, especially several days after the impact, due to their short lifetime (Turco et al., 1982). A bigger particle size would indicate other factors important for NLC formation: the gravity wave amplitude, low temperature, and high water vapor concentration.

Size distribution of particles can be found through a series of ground and space measurements. One of the methods is the polarization analysis that can be efficient if NLC cover the major part of the sky, including large scattering angles. This occurred twice during clear-sky nights in the location of the Wide-Angle Polarization Camera, and larger particles (with the effective size about 100 nm) were recorded both times. Firstly, this fact points to the internal (non-space) causes of NLC formation, and secondly, it explains the high brightness of the clouds. This is also confirmed by satellite data on physical properties of the upper mesosphere. During the first observation night (June 25), the NLC layer was in a cold front of gravity wave with a temperature below 130K, which should cause particle growth (Rusch et al., 2017), and during the second night (June 27), VMR of water vapor increased.

No doubt that the burst-like increase in the meteor smoke concentration related to the meteor activity is a major fact determining NLC occurrence (if the temperature is low enough), but in the case of regular meteor showers or small meteorites, this relation is expected to manifest itself in about a day.

**Acknowledgments**

Authors are thankful to Artem Feofilov (Laboratory of Dynamic Meteorology, Ecole Polytechnique, Paris, France) and Axel Gabriel (Leibniz Institute of Atmosphere Physics, Kühlungsborn, Germany) for useful remarks. The work is supported by Russian Foundation for Basic Research, grant 16-05-00170.